# A study of transition metal K-edge x-ray absorption spectra of $LaBO_3$ (B=Mn, Fe, Co, Ni), $La_2CuO_4$ and $SrMnO_3$ using partial density of states

S. K. Pandey[1], S. Khalid[2], and A. V. Pimpale[1]

[1]UGC-DAE Consortium for Scientific Research, University Campus, Khandwa Road, Indore 452 017, India

[2]National Synchrotron Light Source, Brookhaven National Laboratory, Upton, New York – 11973

E-mail: sk_iuc@rediffmail.com and avp@csr.ernet.in

**Abstract**

The transition metal K-edge x-ray absorption near edge structure (XANES) studies have been carried on $LaBO_3$ (B=Mn, Fe, Co, Ni), $La_2CuO_4$ and $SrMnO_3$ compounds. The theoretical spectra have been calculated using transition metal (TM) 4p density of states (DOS) obtained from full-potential LMTO density functional theory. The exchange-correlation functional used in this calculation is taken under local density approximation (LDA). The comparison of experimental spectra with the calculated ones indicates that single-particle transitions under LDA are sufficient to generate all the observed XANES including those which have earlier been attributed to many-body shake-up transitions and core-hole potentials. The present study reveals that all the experimentally observed features are mainly due to distribution in TM 4p DOS influenced by hybridization with other orbitals. Specifically, for $LaMnO_3$, the feature earlier attributed to shake-up process is seen to arise from hybridization of Mn 4p with La 6p and O 2p orbitals; in $La_2CuO_4$ the features attributed to core hole potential correspond to hybridization of Cu 4p with La 6p, La 5d and O 2p orbitals. To see the effect of inhomogeneous electronic charge distribution and on-site Coulomb and exchange interaction (U′) on the XANES of these compounds generalized-gradient approximation and U′ corrections are incorporated in the calculations. These corrections do not generate any new features in the spectra but affect the detailed intensity and positions of some of the features.

**Introduction**

X-ray spectroscopy is a powerful tool for studying the electronic structure of materials. It involves absorption or emission of x-ray photons and consequent electronic transitions in the different eigenstates of the system. These eigenstates being states of an interacting many-body system are not known exactly. Thus understanding the x-ray spectra of a material is a challenging task. Most of the features observed in the x-ray spectra have been associated with processes involving participation of one electron. There are some features which involve two and/or more than two electrons, such as shake-up and shake-off satellites and collective plasmon features [1,2]. Also, the core hole created in x-ray absorption (XA) process gives rise to additional features in the spectra through the well-known final state effects. Edge singularity is the famous phenomenon observed in many semiconductors and metals, which has origin in the exciton states formed by coupling of such holes with valence electrons [3].

XA spectroscopy is of particular interest as it gives information about the ionic state of the absorbing atom and different unoccupied electronic states. As x-ray energies are in the range of core electron binding energies, the XA process necessarily involves a core electron in the system. In the single-particle picture the absorption process corresponds to transition of an electron from a core state with well defined angular momentum to an unoccupied state of appropriate symmetry. Three theoretical models are commonly employed to calculate the XA spectra: (i) band structure (BS), (ii) configuration interaction (CI), and (iii) multiple scattering (MS) [4,5]. These models use different approximations for calculation of electronic states involved in x-ray transitions. The BS model involves only single-electron excitations to interpret the features of XA spectra, whereas CI and MS methods can be used to interpret the features originating from many-electron excitations. In MS calculation many-electron excitations enter through inelastic scattering processes considered in multi-channel formulation [6]. To interpret the XA spectra of transition metal oxides (TMO) under CI formulation one uses on-site d-d Coulomb and exchange interaction $U'$, charge transfer energy $\Delta$, and d-L transfer integral T (L stands for ligand) as parameters of calculation [7]. As CI calculations do not use any self-consistency scheme, the values of parameters used in the calculations are crucial and should be reasonably accurate. BS and MS methods use self-consistency schemes, thus the input parameters for these models have more tolerance.

In general, using first order perturbation theory for interaction of radiation with matter the transition probability per second ($w$) for x-ray absorption is given by Fermi golden rule

$$w = \frac{2\pi}{\hbar} \left| \langle \psi_F | H_{int} | \psi_I \rangle \right|^2 \rho(E_F), \text{ with condition } E_F = E_I + \hbar\omega \qquad (1)$$

where $H_{int} = \frac{e}{mc} \sum_i \vec{p}_i \cdot \vec{A}(\vec{r}_i, t)$ is the interaction Hamiltonian for the process of XA and $\psi_I$, $\psi_F$, and $\rho(E_F)$ are initial eigenstate, final eigenstate and density of final states of the interacting many-body system, respectively, $E_I$ and $E_F$ are the corresponding eigenvalues, $\vec{A}(\vec{r}, t)$ represents the vector potential for radiation and $\hbar\omega$ is the x-ray photon energy. The summation in $H_{int}$ runs over all the electrons in the system. The eigenstate of interacting many-body system can always be written using different combinations of single-particle eigenstates. Since $H_{int}$ is a sum over single-particle operators, in matrix element $\langle \psi_F | H_{int} | \psi_I \rangle$ only one single-particle state in $\psi_I$ will contribute at a time. Therefore, under first order perturbation theory one cannot get features associated with involvement of more than one electron in the absorption process.

To find $\psi_I$, $\psi_F$, and $\rho(E_F)$ different approximation methods are used. In the CI calculations $\psi_I$ and $\psi_F$ are formed by taking the linear combination of appropriate single-particle wave functions corresponding to the different configurations. In MS calculations one considers the absorption process as arising from a single electron transition from core state $\phi_i$ to a state in the continuum $\phi_s$; all other electrons are considered as passive, and their overlap is taken as a factor, ideally unity, but in practice between 0.7 to 0.9. $\phi_s$ is calculated by summing over all the scattered states obtained by considering the different scattering processes, from one atomic scatterer, two atomic scatterers and so on. In the BS approach one uses the single-particle density of states of appropriate symmetry and a matrix element for corresponding transition involving single-particle states with the rest of electrons treated as passive. In this work we employ BS calculations using density functional theory (DFT) [8,9] to calculate single-particle density of states (DOS) of a periodic solid and use the same to generate XA spectra. DFT uses self-consistent equations

$$\left[-\frac{1}{2}\nabla^2 + v_{eff}(\vec{r})\right]\psi_i(\vec{r}) = \varepsilon_i\psi_i(\vec{r}) \qquad (2)$$

where $v_{eff}(\vec{r}) = v(\vec{r}) + \int \frac{n(\vec{r})}{|\vec{r}-\vec{r}'|}d\vec{r}' + v_{xc}(\vec{r})$ \qquad (3)

and $n(\vec{r}) = \sum_{i=1}^{N}|\psi_i(\vec{r})|^2$ \qquad (4)

where N stands for total number of electrons

The accuracy of single-particle states and their DOS obtained from DFT depends on the accuracy of the exchange-correlation potential $v_{xc}(\vec{r})$ used, which is taken under local density approximation (LDA) [9,10] valid for systems with smooth electronic charge distribution. For simple systems where electron densities vary slowly, LDA has been used successfully for understanding the XA spectra [11-13]. The rare earth transition metal oxides under study here- $LaMnO_3$, $SrMnO_3$, $LaFeO_3$, $LaCoO_3$, $LaNiO_3$, and $La_2CuO_4$- have been known to display many-body and core hole potential features in their transition metal [TM] K-edge x-ray absorption near edge structures (XANES). Earlier, XANES of these compounds have been analysed by using CI and MS calculations. Most of the work has been concentrated on $LaMnO_3$, $LaNiO_3$, and $La_2CuO_4$ [14-25] and very little work is found in the literature on $SrMnO_3$, $LaFeO_3$, and $LaCoO_3$ [26,27]. These studies indicate that although most of the near edge structures can be attributed to single-particle transitions, there are some features that cannot be accounted for without considering many-body and core hole effects. For instance, Iganatov et al. [16] have attributed the experimentally observed feature at about 6 eV above the main absorption peak in $LaMnO_3$ to many-body shake-up transition. Similarly, some observed features in $La_2CuO_4$ have been attributed to fully relaxed and partially relaxed core hole states [20,21]. Moreover, Some features observed in $LaNiO_3$ and $La_2CuO_4$ have been attributed to mixed configurations and mixed valence states [18,20].

In the band structure scheme using DFT one cannot directly account the many-body features, core hole and mixed valence effect etc. For the compounds under study electron density variation is appreciable and LDA may not be applicable in interpreting the XA spectra. One may need higher order of approximations like generalized-gradient approximation (GGA), which takes care of spatial variation of electron density [9]. Ravindran et al. [28] have earlier emphasized the importance of GGA correction and spin orbit coupling in the theoretical calculations of $LaMnO_3$ to interpret the spectral features of this compound. These compounds fall in the category of highly correlated systems where U′ plays an important role in deciding the electronic states of the system and one expects U to influence the XA spectra. The electronic structure of $La_{1-x}Ca_xMnO_3$ has been studied by Qian et al. [29] using spin polarized XAS and the data were interpreted using LDA and LDA+U BS calculations. The importance of electron correlations in understanding the spectra in the BS formulation has been noted in this work although the experimental spectra were not calculated explicitly using BS. These systems are believed to be charge transfer type insulators and transfer of electrons from ligand to metal is common. The presence of core hole created in x-ray transitions affects such transfer and can produce additional features in the XA spectra. Thus one expects some features in the XA spectra with origin beyond one electron process, which cannot be accounted for in first order perturbation calculations using LDA. Hence, a priori, one expects the failure of LDA in interpreting all the experimentally observed features in XA spectra of these compounds.

We have recently shown that for $LaCoO_3$, Co K-edge XANES can be explained by convoluting Co 4p DOS obtained by full-potential (FP)- LMTO DFT calculations and taking into consideration the hybridization of different participating orbitals with reasonable density at appropriate energy [27]. These calculations were done using LDA+U and taking GGA into account, since it is known that the ground state character of the system is not generated in LDA alone. In this calculation we noted that the main features of the Co 4p DOS, being well away from the Fermi level, are not much affected by the U′ and GGA corrections. The accuracy of DOS at energies much above the Fermi level is known to be not so good in LMTO calculations. However, use of FP-LMTO increases the accuracy. It has been successfully used to interpret all the observed features in the experimental optical spectra up to about 30 eV above the Fermi level [28,30]. Thus such calculations may also be used for interpretation of XANES. We, therefore, thought it worthwhile to calculate the TM 4p DOS using FP-LMTO DFT under LDA for other TM oxides showing XA features beyond single-particle excitations. In addition to convoluting the TM 4p DOS to simulate the lifetime broadening effect, we incorporated the matrix element and computed the XANES. The effect of $U'$ and GGA corrections are also computed.

Here we report the TM K-edge XA studies of $LaMnO_3$, $SrMnO_3$, $LaFeO_3$, $LaCoO_3$, $LaNiO_3$, and $La_2CuO_4$ compounds by comparing the experimental data with BS calculations under DFT formulation using LDA and also GGA and $U'$ corrections. For $LaNiO_3$, and $La_2CuO_4$ we have used the data of Sahiner et al. and Alp et al., respectively [17,20]. For $LaFeO_3$, unfortunately good quality digitizable data were not available. For remaining compounds- $LaMnO_3$, $SrMnO_3$, and $LaCoO_3$ we have used our own data. The XA spectra have been calculated using the TM 4p DOS obtained from BS calculations and matrix elements from atomic like core state and final state as a confluent hypergeometric function. Contrary to earlier reports on these compounds where some of the near edge features have been attributed to many-body transitions and core hole effects, this study reveals that all the experimentally observed features are generated using these calculations within LDA indicating that single-particle approximation is sufficient to understand the XA spectra. The origin of different features has been identified based on the hybridization of different orbitals with the TM 4p orbital. The effect of GGA and $U'$ corrections on XA spectra has been discussed using LDA+GGA and LDA+U calculations. This study indicates that these corrections are not strong enough to generate any new feature in the XA spectra but affect the details of positions and intensities of some of these features.

**Experimental and computational details**

Single-phase polycrystalline powder samples of $LaMnO_3$, $SrMnO_3$ and $LaCoO_3$ were prepared by standard solid-state route and combustion method [31,32]. The samples were characterized by powder x-ray diffraction (XRD) technique. Lattice parameters obtained from Rietveld analysis matched well with literature. Room temperature Mn and Co K-edges XAS data were recorded at beamline X-18 B at the National Synchrotron Light Source, Brookhaven National Laboratory. The storage ring was operated at 2.8

GeV and 300 mA. The beamline used a Si (111) channel cut monochromator. The horizontal acceptance angle of the beam at the monochromator was 1 mrad. The vertical slit size used in this experiment was 1 mm, corresponding to an energy resolution of about 0.8 eV at the Mn K-edge. The average photon flux for this bandwidth was $10^{10}$ photons/sec. The monochromator was detuned by 35 % to reduce the higher harmonics. The incident ($I_0$) and the transmitted beam ($I_t$) were measured by sealed ion chambers, with a combination of gases for appropriate absorption. Standard TM foil was placed between the detectors $I_t$ and $I_{ref}$ for energy reference and to check the stability of the beamline and optical system. The sample powder sieved through a 400 mesh and again grinded with mortar and pestle was spread uniformly on a cellophane tape and four layers of this tape were used to minimize the pinhole and brick effects.

Spin polarized LDA, LDA+GGA, and LDA+U band structure calculations were carried out using LMTART 6.61 [33]. For calculating charge density, FP-LMTO method working in plane wave representation was used. The muffin-tin radii used in the calculations for each compound are listed in table 1. The charge density and effective potential were expanded in spherical harmonics up to $l = 6$ inside the sphere and in a Fourier series in the interstitial region. The initial basis set included 6s, 6p, 5d, and 4f valence and 5s, 5p semicore orbitals of La; 5s, 5p, and 4d valence and 4s, 4p semicore orbitals of Sr; 4s, 4p, and 3d valence and 3p semicore orbitals of transition metals and 2s and 2p valence orbitals of O. The two-kappa basis for these orbitals is included in the calculations. The system contains heavy ions where the relativistic effects are important for core orbitals. The relativistic effects have also been included in the calculations. The exchange correlation functional of the density functional theory was taken after Vosko, Wilk, and Nussair [10] and the generalized-gradient approximation scheme of Perdew et al. [34] was also invoked. In the LDA+U method three input parameters are required for 3d electrons. They are Slater integrals $F^0$, $F^2$, and $F^4$. These integrals are directly related with on-site Coulomb interaction (U) and exchange interaction (J) by relations $U = F^0$, $J = (F^2 + F^4) / 14$, and $F^4/ F^2 \sim 0.625$. We have used U = 6.0 eV and J = 0.95 eV for all compounds. These values are closer to those reported by Yang et al. [35]. (6, 6, 6) divisions of the Brillouin zone along three directions for the tetrahedron integration were used to calculate the DOS. Self-consistency was achieved by demanding the convergence of the total energy to be smaller than $10^{-5}$ Ryd/cell. In each case the energy origin is the Fermi level.

Under dipole approximation the K-edge absorption process occurs due to transition of electron from s-symmetric state to p-symmetric unoccupied states. Due to participation of localized states in transition the contribution of states of atoms other than the TM to the TM K-edge absorption is negligible. Thus TM K-edge absorption can be identified to the electronic transition from TM 1s to 4p states. To calculate the absorption spectra we need matrix elements and DOS of TM 4p states. We have calculated the matrix element $\langle \psi_F | H_{int} | \psi_I \rangle$ by using $|\psi_I\rangle$ as the 1s hydrogenic wave function and $|\psi_F\rangle$ as confluent hypergeometric function (CHF) [36]. Matrix element thus obtained has been used to obtain the transition probability per second ($w$) by using Eq. (1). The modulus of the matrix element is a monotonic function. It would have been more appropriate to use the final state wave function as given by DFT rather than the CHF in the calculations of matrix element. However, it is not possible to extract such function from the LMTART program. Use of CHF is known to be reasonable for high-energy scattering states; although it is somewhat questionable in the vicinity of the main-edge. To calculate the absorption spectrum, calculated $w$ is convoluted with a Lorentzian having FWHM of 2.1 eV to account for the lifetime of TM 1s core hole generated during the absorption process [37]. The experimental spectrum has been rigidly shifted to match the position of its most intense peak with that of calculated one. The calculated spectrum is normalized to the same intensity as that of this most intense peak. In the various figures the calculated and experimental spectra are shifted along y-axis for the sake of clarity.

**Results and discussion**

The experimental absorption spectra of all the compounds can be divided into three regions: (1) pre-edge, (2) main-edge, and (3) post-edge. In the pre-edge region different compounds show different pre-edge structures. These structures are generally attributed to quadrupole allowed 1s to 3d transitions [27,38,39]. We discuss below the main-edge and post-edge structures of each compound separately using LDA for interpretation. The effect of GGA and U´ corrections on the XA spectra is discussed at the end of the section

The origin of different structures observed in the experimental XA spectrum can be explained by considering the partial DOS of different symmetry in the corresponding energy range. The DOS is sensitive to chemical bonding and the crystal structure. Partial DOS corresponding to a particular orbital used in band structure calculations as one of the basis states is relevant for XA as the core state involved is characterized by a well defined angular momentum. In the case of TM K-edge the partial DOS of relevance arises from TM 4p orbital. The other basis state functions at approximately the same energy as the TM orbital modify the partial DOS in different compounds. It is these modifications that give rise to different features in the K-edge XANES of different compounds. The changes in the partial DOS can be understood in terms of the relative weightage of the TM 4p orbital in the basis states giving rise to the energy band at the particular energy. Higher the component of the empty TM 4p orbital more would be the XA corresponding to TM 1s to 4p transition. The TM 4p DOS at a particular energy and other partial DOS at this energy imply the relative weightage of these states in the final self-consistent band structure. These hybridizing orbitals influence the 4p DOS and generate the observed experimental features. Higher the partial DOS of a orbital more would be its effect on the TM 4p DOS. The calculated and experimental spectra for different TM oxides are discussed below in the light of above consideration.

### $LaMnO_3$ and $SrMnO_3$

The experimental and calculated Mn K-edge spectra of $LaMnO_3$ are shown in the upper panel of figure 1 by open circles and dotted lines, respectively. It is evident from the figure that there are five structures in the experimental spectrum denoted by A, B, C, D, and E. Although structure C around 20 eV is not prominent, its presence is confirmed by taking the second derivative of the spectrum which shows a dip around this energy. All the structures present in the experimental spectrum are seen in the calculated one and denoted by arrows. The energy position of all the features match well except peak A which is at slightly higher energy in the calculated spectrum. Such a deviation is expected as the energy positions of the near edge features are influenced by the effect of the core hole potential. A noteworthy result is the creation of feature C in the calculated spectrum that was absent in the single-particle MS calculations carried out by Ignatov et al. [16]. They have attributed this feature to many-body shake-up transition.

The partial DOS of La 6p, Mn 4p, and O 2p sates are plotted in the lower panel of figure 1, partial DOS of rest of the states in the energy range of our interest is not appreciable. In the energy range of 8-14 eV, where feature A is observed, the La 6p DOS is about 4.7 times more than that of O 2p. Therefore, this feature can be attributed to hybridization effect of Mn 4p orbital with that of La 6p and it should be more sensitive to La site than O site. Hence, any replacement of La atoms will affect the intensity of peak A. As La is replaced, hybridization of Mn 4p state with La 6p is reduced and relative contribution of Mn 4p increases thereby increasing the intensity of absorption peak at this energy. This is also experimentally seen in the systems with La atoms replaced by Ca or Sr atoms [16,31]. This result is not in conformity with that of Ignatov et al. [16], which they have attributed to the hybridization of Mn 4p and La 5d states. According to our calculations the energy position of La 5d band comes out to be around 5.5 eV, which is well below the position of Mn 4p band contributing to peak A. After peak A one can see a steep rise in the calculated spectrum up to feature B and then decrease through feature C and beyond up to feature D. Mn 4p DOS in the energy range 13-17 eV will contribute to the region up to feature B. In this energy range ratio of La 6p DOS to O 2p DOS is about 5, therefore in this region also the hybridization with La 6p orbital is more important than that with O 2p. Mn 4p DOS in the energy region of 18-22.5 eV will contribute to structure C in the post-edge region. In this region La 6p DOS is only about 2.6 times greater than that of O 2p DOS; hence the effect of O 2p state on this feature is more in comparison to structures A and B. To see the experimental verification of this behaviour one can vary the oxygen stoichiometry and compare the change in relative intensity of features B and C. Subías et al. [40] have carried out Mn K-edge XA experiments on oxygen stoichiometric and non- stoichiometric $LaMnO_3$ compounds. A closer look at the XA spectra given in figure 1 of ref. [40] reveals that with increase in oxygen content feature C becomes less and less prominent, in accordance with our expectations. DOS of all the states are almost comparable in the energy range of structure E. Thus, the effect of hybridization of all the states on Mn 4p will be almost the same.

The experimental and calculated Mn K-edge spectra of $SrMnO_3$ are shown in the upper panel of figure 2. Experimental spectrum of this compound also shows structures A, B, C, D, and E. Structures A and C are more prominent in $SrMnO_3$ in comparison to $LaMnO_3$. In the calculated spectrum these structures are present. Feature C is also more prominent in the calculated spectrum of this compound. The partial DOS of Sr 5p, Mn 4p, and O 2p states are shown in the lower panel of figure 2. It is evident from the figure that in the energy range of feature A, Sr 4d and 5p DOS are contributing more and therefore hybridization of these orbitals with Mn 4p is crucial for structure A. In the region of structure B, Sr 5p and O 2p DOS are almost the same whereas in region of feature C, Sr 5p DOS has more contribution. The amount of O 2p DOS in the energy region of feature C is almost the same for both the compounds. But the DOS of Sr 5p is less in comparison to La 6p. Therefore, one expects decrease in the hybridization effect and hence more DOS of Mn 4p character in the energy region of feature C. The increase in Mn 4p DOS is seen in the calculation and this is corresponding to the increased prominence of feature C in $SrMnO_3$ as compared to that in $LaMnO_3$.

## $LaFeO_3$

The calculated Fe K-edge spectrum of $LaFeO_3$ is shown in the upper panel of figure 3. Unfortunately we have no access to experimental data in this case; only the Fe K-edge data of Wu et al. [26] was noted in the literature and interested reader can refer it. The experimental spectrum shows all the pronounced features seen in the calculated spectrum; however, the latter contains additional features which are not seen in the data. This may be due to instrumental factors. The calculated spectrum shows similar structures as seen in the case of $LaMnO_3$, although the structure C in $LaFeO_3$ is less pronounced. Similar to $LaMnO_3$, structure A has origin in the hybridization of Fe 4p orbital with La 6p orbital. For structure B hybridization of both La 6p and O 2p orbitals with Fe 4p seems to be important as the DOS of La 6p and O 2p are nearly the same in the energy range contributing to this peak. The partial DOS of La 6p, Fe 4p, and O 2p states are shown by solid, dotted, and dashed lines, respectively in lower panel of figure 2. This figure clearly indicates that Fe 4p DOS around 20 eV contributes to the feature C. In contrast to $LaMnO_3$, in the energy range of feature C La 6p DOS is about 5.4 times than that of O 2p, therefore any change on La site will have pronounced effect on this feature. This observed difference in two compounds seems to be due to lesser spread of Fe 4p DOS in the main-edge region as compared to the corresponding Mn 4p DOS. Moreover, in the energy range of structures E and F the influence of O 2p orbital is more pronounced.

## LaCoO$_3$ and LaNiO$_3$

The Co and Ni K-edges experimental spectra of LaCoO$_3$ and LaNiO$_3$ are shown by open circles in the upper panels of figures 4 and 5, respectively. Both spectra show six structures- two on the main-edge region and the other four in the post-edge region. The calculated spectra of both the compounds shown by dotted lines in the upper panels of figures 4 and 5 also show all the six structures. Although feature C is not pronounced in the experimental spectra, it is clearly visible in the calculated ones. The reason for the non-observance of feature C in the experimental spectra is that the instrumental broadening masks it. On including the instrumental broadening in calculations, feature C becomes feeble. The instrumental-broadened spectrum of LaCoO$_3$ is shown by solid line in the upper panel of figure 4. The small features seen in post-edge region of lifetime broadened spectrum are also absent in the instrumental broadened spectrum. Therefore, spectrum indicated by the solid line is a better representative of the experimental spectrum.

The calculated partial DOS of La 6p, Co 4p and O 2p states for LaCoO$_3$ and partial DOS of La 6p, Ni 4p and O 2p states for LaNiO$_3$ are plotted in the lower panels of figures 4 and 5, respectively. It is clear from the figures that in the energy region of feature A La 6p DOS has major contribution. Thus hybridization of La 6p with Co (Ni) 4p state contributes most to this feature as in the case of LaMnO$_3$ and LaFeO$_3$. In the energy region of structure B there is not much difference in La 6p, Co (Ni) 4p and O 2p DOS indicating the highly mixed character of these states. Hence, similar to LaFeO$_3$, the contributions of La 6p and O 2p states to the main feature are about the same. However, in the region of feature C the La 6p DOS is about 3 times that of O 2p for both LaCoO$_3$ and LaNiO$_3$, which is slightly more than in LaMnO$_3$. Since the contribution of La 6p and O 2p states to the feature B is almost the same, one expects more influence of oxygen non-stoichiometry on this feature than feature C. Such behaviour is opposite to that in LaMnO$_3$, where the effect of oxygen non-stoichiometry was more on the feature C. In the region of structure E, partial DOS of all the states are almost equal, whereas in the energy region of structure F contribution of La 6p state is negligible. Thus, structure E can be attributed to the hybridized states of La 6p, Co (Ni) 4p and O 2p, similar to structure B. However, structure F is attributed to the hybridization of Co (Ni) 4p state with that of O 2p. Here it is interesting to compare the MS scattering calculations of García et al. [19] using an X-α approximation for exchange correlation potential with our band structure calculations. According to their calculation structure F of LaNiO$_3$ has major contribution from the $3d^8\underline{L}$ configuration. This indicates the mixing of Ni 3d orbital with O 2p orbital. Our observation of large O 2p unoccupied DOS is in conformity with transfer of O 2p electron to the TM.

## La$_2$CuO$_4$

The Cu K-edge experimental spectrum of La$_2$CuO$_4$ is shown in the upper panel of figure 6 by open circles. This spectrum shows seven structures- two below the main feature C and four above it. Features A and B are respectively about 10 and 4.5 eV below the main feature C and features D, E, F, and G are about 5, 7, 11, and 14 eV, respectively above it. The calculated spectrum indicated by dotted line has all the features seen in the experimental spectrum. For better matching of the edge shape we have convoluted the lifetime broadened spectrum with a gaussian having FWHM of 3 eV to account for the instrumental and other broadening effects. The energy positions of all the structures in the calculated spectrum match well with the corresponding experimental ones. This is interesting, because we have seen small deviations in the energy positions of some features in above perovskite type compounds. Such deviations are usually attributed to the effect of core hole potential. This indicates that the effect of core hole potential on the K-edge x-ray spectra of layered compounds like La$_2$CuO$_4$ may not be as important as in the case of perovskite type compounds.

To identify the contribution of different states to the observed structures we have plotted the partial DOS of La 6p, La 5d, Cu 4p, and O 2p as the DOS of these states are significant in the energy range of interest. These DOS are plotted in the lower panel of figure 6. It is evident from the figure that feature A can be attributed to the transition to Cu 4p states hybridized with La 5d and O 2p states. In region of structure B, La 6p state also contributes, however, the effect of La 5d states is more on this feature. With increase in energy the La 6p DOS increases up to 20.7 eV. In the energy region of main peak C, the contribution of La 6p and O 2p is significant with relatively less contribution from La 5d state. Therefore, structure C can be attributed to the strongly hybridized Cu 4p state with that of La 6p and O 2p. In the case of the post edge structures D and E, the effect of La 5d state is more with a relatively small contribution of O 2p DOS; therefore, these features can be attributed to hybridized Cu 4p and La 5d states. Similarly, in the region of structures F and G the La 5d and O 2p DOS are negligible in comparison to La 6p DOS. Hence these features can be identified with the hybridization of Cu 4p state with La 6p.

A comparison of results obtained from this calculation with other calculations will be quite interesting. To interpret the Cu K-edge XA spectrum of $La_2CuO_4$ Alp et al. [20] have carried out single-particle calculation using self-consistent-field theory and discrete variational method. This calculation is unable to produce feature E and hence by comparing the energy position of main peak of $KCuO_2$ they attributed it to $Cu^{3+}$ configuration. However, Kosugi et al. [22,41] have attributed features A and B to 1s- 4p$\pi$ transitions to well screened (through ligand to metal charge transfer) and poorly screened core hole states, respectively based on the MS calculations. Similarly, features C and E have been attributed to 1s- 4p$\sigma$ transitions corresponding to well screened and poorly screened core hole states, respectively. Therefore, according to their results these features are related with core holes and their screening. But our calculations show that these structures arise without the core hole and solely from distributions of Cu 4p DOS in presence of La 5d, La 6p, and O 2p states. This is quite interesting, as the importance of core hole potential has been emphasized by many workers. Our study indicates that core hole potential is not strong enough to change the final state in such a way as to create a new feature in XA spectrum, however the energy positions and intensities of the features may be affected by the core hole potential.

**Influence of non-uniform electronic charge distribution and d-d Coulomb and exchange interaction (U′) on x-ray absorption spectra**

Normally it is believed that these compounds have non-uniform electronic charge distribution due to localized and extended nature of transition metal 3d and O 2p states, respectively. Moreover, the importance of U′ for these compounds has been emphasized in interpreting the XA spectra by some workers as remarked earlier. By invoking

LDA+GGA and LDA+U in the calculations one can see the effect of non-uniform electronic charge distribution and U′, respectively. We have calculated the K-edge XA spectra of all the compounds by using the TM 4p DOS obtained from LDA, LDA+GGA, and LDA+U separately. All the spectra have been normalized to unity at the position of main feature. Moreover, for comparing these effects the different spectra obtained from LDA+GGA and LDA+U were shifted to match the position of the main peak with that obtained from LDA.

The calculated XA spectra of $LaMnO_3$, $LaFeO_3$, $LaCoO_3$, and $La_2CuO_4$ are plotted in the upper and lower panels of figures 7 and 8. Solid, dashed, and dotted lines indicate the spectra obtained from LDA, LDA+GGA, and LDA+U calculations, respectively. It is evident from the figures that consideration of LDA+GGA and LDA+U does not create any new structures. Thus, these corrections are not strong enough to generate any extra features in the spectra. This was expected because LDA was able to generate all the experimentally observed features as discussed above. However, the position and intensity of the various features are sensitive to these corrections. In the case of $LaMnO_3$, GGA correction seems to affect the spectrum considerably by shifting the relative positions of the features and broadening the spectrum in the edge region. Intensity of structure E also changes drastically indicating that any redistribution of electronic changes due to doping and/or structural changes will directly affect this feature. The U′ correction slightly increases the intensity in the post edge region keeping overall spectral shape almost the same. Surprisingly, the effect of GGA and U′ corrections on the XA spectra of TM K-edge in $LaFeO_3$, $LaCoO_3$, and $LaNiO_3$ is least. These corrections slightly change the intensity of the post edge structures. However, one can see the clear effect of GGA correction on feature C in $LaCoO_3$ and $LaNiO_3$ where its intensity decreases. This is in contrast to $LaMnO_3$ where intensity of feature C increases on invoking GGA correction. Moreover, the effect of these corrections is quite considerable on the XA spectrum of $La_2CuO_4$ where all the structures are affected. As in the case of $LaMnO_3$, GGA broadens the edge of $La_2CuO_4$ and changes the relative positions of different structures. The effect of U′ correction is more on structure G where its intensity increases significantly.

Finally, it may be noted that for all the calculated and corresponding experimental spectra, the intensity falls off much sharply in the calculated spectra as compared to the experimental data. The calculations were calibrated so as to match at the main peak. Thus the decline in intensity on the high-energy side for the calculated spectra may be attributed to the fact that the final state wave function used for matrix element calculations was not a solid state wave function. Moreover, the absolute intensity of absorption spectrum depends on the effects like inelastic losses, extrinsic losses etc. It may be mentioned that such disagreement in intensity in experiment and calculation is quite known in literature and Rehr and Albers [4] have commented on it as one of the challenging problems in the field. We want to comment on the spin-orbit (SO) interaction known to affect the multiplet states which are crucial for understanding the XANES features through multiplet calculations. Our calculations indicate that the effect of SO interaction decreases for high-energy states in the continuum. For the final states involved in the x-ray absorption process this effect seems to be not so important, and, as a consequence DOS corresponding to these states remain almost intact. This may be because of the quenching of the orbital angular momentum for these states.

**Summary and concluding remarks**

The transition metal (TM) K-edge x-ray absorption (XA) studies have been carried out on $LaMnO_3$, $SrMnO_3$, $LaFeO_3$, $LaCoO_3$, $LaNiO_3$, and $La_2CuO_4$ compounds. The theoretical TM K-edge spectra of all the compounds have been calculated using TM 4p density of states obtained from band structure calculations using density functional theory and considering exchange correlation energy under local density approximation (LDA). The matrix elements used in these calculations have been obtained by considering 1s hydrogenic wave function as initial state and confluent hypergeometric function as final state. The comparison of experimental spectra with the calculated ones reveals that the single-particle calculations under LDA are sufficient to generate all the experimentally observed features. This result is quite interesting in the light of earlier studies on $LaMnO_3$ and $La_2CuO_4$ where some of the near edge features have been attributed to many-electron processes and to states due to core hole potential. In an earlier report on $LaMnO_3$ where the feature about 6 eV above the main peak has been attributed to many-body shake-up peak, the present study reveals that this peak has origin in the distribution of Mn 4p DOS due to hybridization of Mn 4p state with the La 6p and O 2p states. The absence of many-body features in the TM K-edge spectra is not surprising as the photon energy involved in the transition is very high and any direct coupling of radiation with many body excitations like plasmon and exciton would be absent, although these excitations could couple with radiation in the higher order approximation. Further the simultaneous coupling of core hole and the photoelectron produced in the absorption process with the collective excitation of the valence electron system tend to cancel each other out [42]. Similarly, four near edge features- three on the edge and one just after the main peak- observed in $La_2CuO_4$ and attributed to transition to states in the presence of fully relaxed and partially relaxed core hole are also present in our calculated spectrum and attributed to transition to Cu 4p states hybridized with La 5d, La 6p, and O 2p states. This study reveals that the effect of core hole is not strong enough to create any extra features in the XA spectrum. The correction due to core hole may of course improve the intensity and position of some features. The inclusion of generalized-gradient correction in the LDA calculations to account for inhomogeneous electronic charge distribution indicates that the effect of this correction is also not strong enough to generate new feature in the XA spectra. However, this correction is required to see the effect on the positions and intensities of the observed features, especially in $LaMnO_3$ and $La_2CuO_4$. Similarly, the inclusion of on-site Coulomb and exchange correction ($U'$) for the TM 3d orbitals under LDA+U formulation has only a small effect on the intensity and position of the observed features in the XA spectra, indicating that the interpretation of XA spectral features using $U'$ is not required.

**Acknowledgements**

The authors would like to thank Ashwani Kumar for his help in developing a program to calculate matrix element. SKP thanks UGC-DAE CSR for financial support.

**Figure captions**

Figure 1   In the upper panel the experimental and calculated Mn K-edge x-ray absorption spectra of $LaMnO_3$ are indicated by open circles and dotted line, respectively. In the lower panel La 6p, Mn 4p, and O 2p density of states per formula unit are denoted by solid, dotted, and dashed lines, respectively.

Figure 2   In the upper panel the experimental and calculated Mn K-edge x-ray absorption spectra of $SrMnO_3$ are indicated by open circles and dotted line, respectively. In the lower panel Sr 5p, Sr 4d, Mn 4p, and O 2p density of states per formula unit are denoted by solid, dash dotted, dotted, and dashed lines, respectively.

Figure 3   In the upper panel the calculated Fe K-edge x-ray absorption spectrum of $LaFeO_3$ is denoted by dotted line. In the lower panel La 6p, Fe 4p, and O 2p density of states per formula unit are denoted by solid, dotted, and dashed lines, respectively.

Figure 4   In the upper panel the experimental and calculated Co K-edge x-ray absorption spectra of $LaCoO_3$ are indicated by open circles and dotted line, respectively. The solid line represents the calculated spectrum when instrumental broadening is also included. In the lower panel La 6p, Co 4p, and O 2p density of states per formula unit are denoted by solid, dotted, and dashed lines, respectively.

Figure 5   In the upper panel the experimental [17] and calculated Ni K-edge x-ray absorption spectra of $LaNiO_3$ are indicated by open circles and dotted line, respectively. In the lower panel La 6p, Ni 4p, and O 2p density of states per formula unit are denoted by solid, dotted, and dashed lines, respectively.

Figure 6   In the upper panel the experimental [20] and calculated Cu K-edge x-ray absorption spectra of $La_2CuO_4$ are indicated by open circles and dotted line, respectively. In the lower panel La 6p, La 5d, Cu 4p, and O 2p density of states per formula unit are denoted by solid, dash dotted, dotted, and dashed lines, respectively.

Figure 7   The calculated Mn and Fe K-edges x-ray absorption spectra of $LaMnO_3$ and $LaFeO_3$ are given in upper and lower panels, respectively. The solid, dash, and dotted lines represent the calculated spectra under LDA, LDA+GGA, and LDA+U schemes, respectively.

Figure 8   The calculated Co and Cu K-edges x-ray absorption spectra of $LaCoO_3$ and $La_2CuO_4$ are given in upper and lower panels, respectively. The solid, dash, and dotted lines represent the calculated spectra under LDA, LDA+GGA, and LDA+U schemes, respectively.

**Table caption**

Table 1   The muffin-tin radii of La, Sr, Mn, Fe, Co, Ni, Cu, and O atoms in atomic units used in calculations for different compounds.

Table 1

| Samples | La/Sr | M | O |
|---|---|---|---|
| $LaMnO_3$ | 3.401 | 1.981 | 1.662 |
| $SrMnO_3$ | 3.439 | 1.986 | 1.597 |
| $LaFeO_3$ | 3.576 | 2.041 | 1.674 |
| $LaCoO_3$ | 3.509 | 2.001 | 1.637 |
| $LaNiO_3$ | 3.526 | 2.011 | 1.646 |
| $La_2CuO_4$ | 2.724 | 1.930 | 1.644, 1.741 |

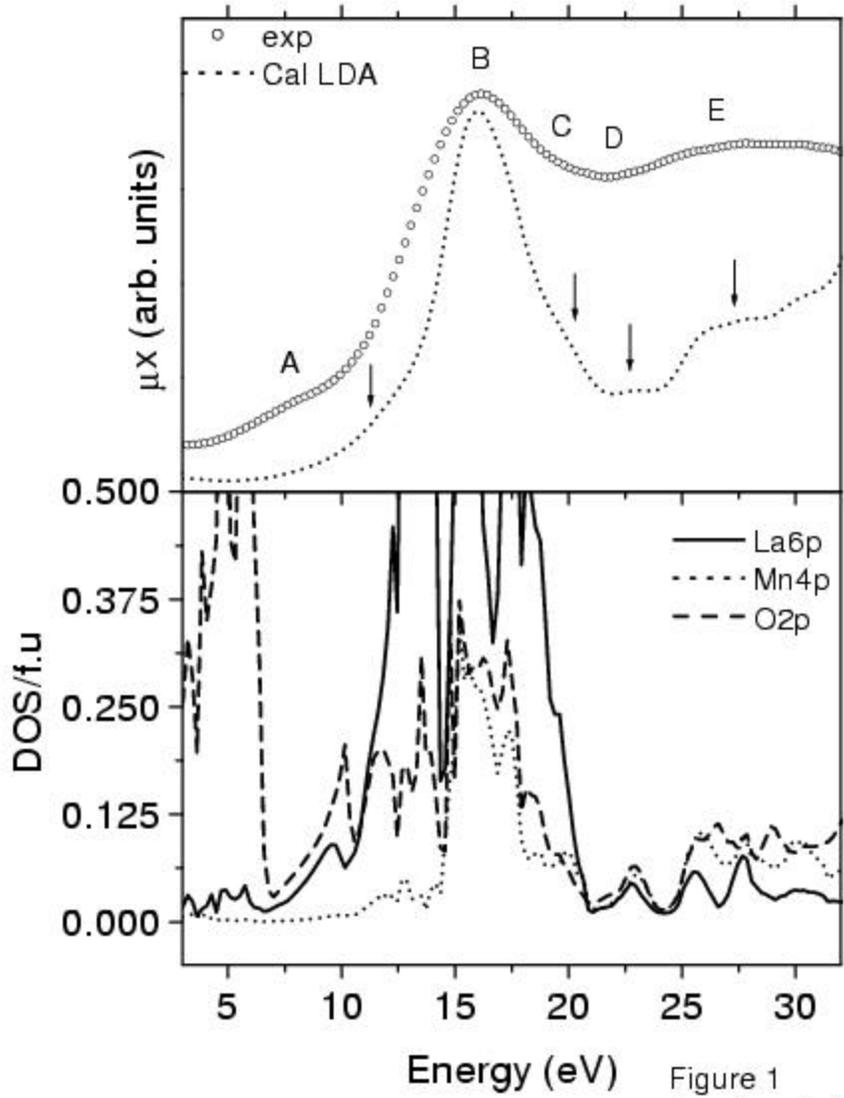

Figure 1
Pandey et al.

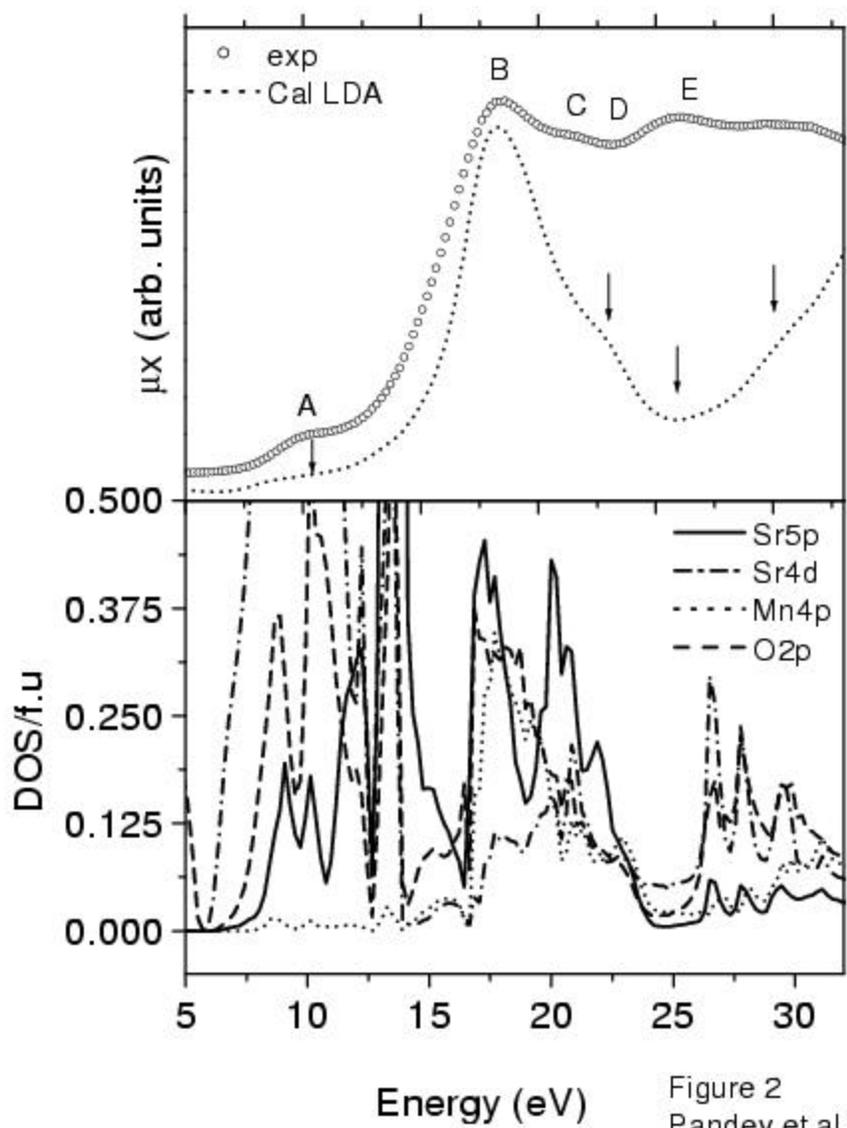

Figure 2
Pandey et al.

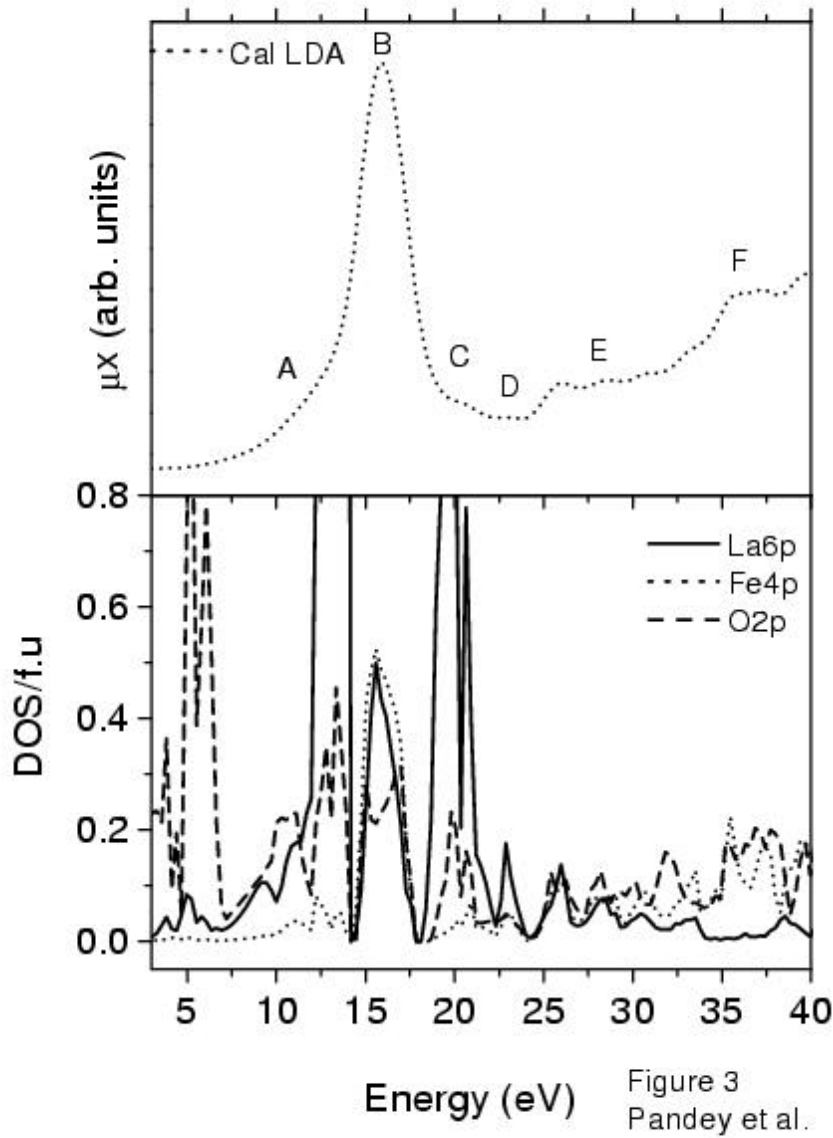

Figure 3
Pandey et al.

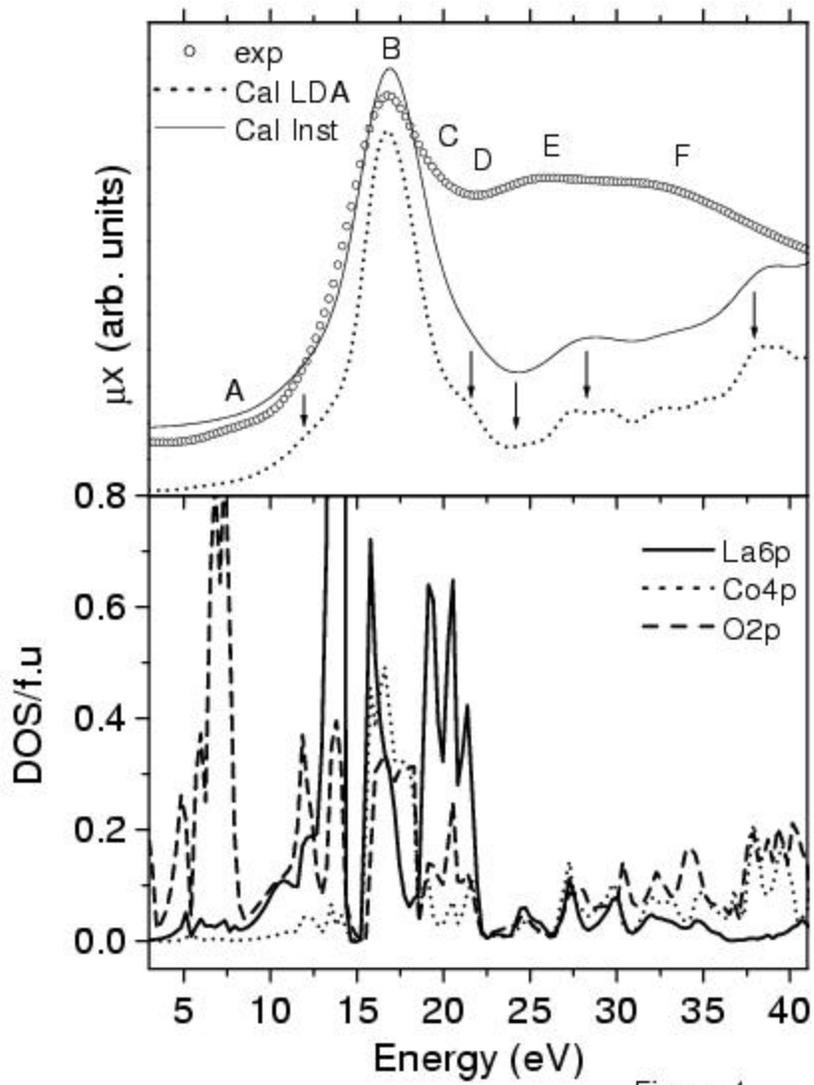

Figure 4
Pandey et al.

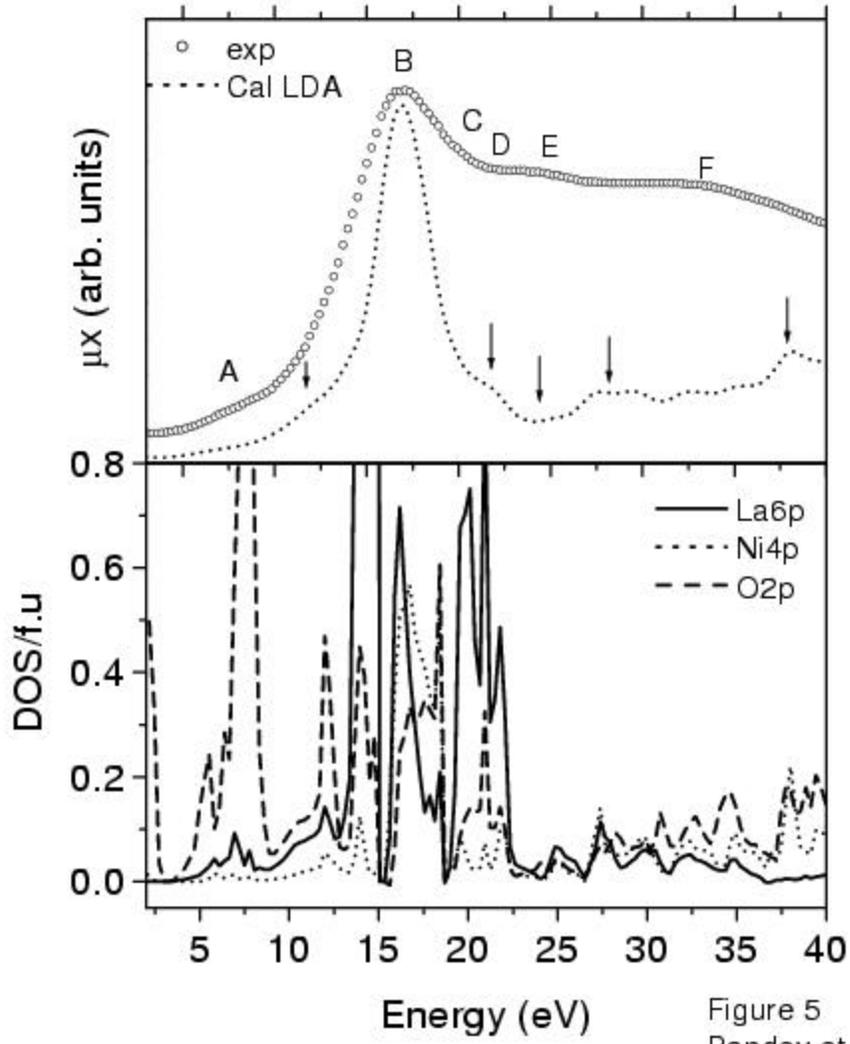

Figure 5
Pandey et al.

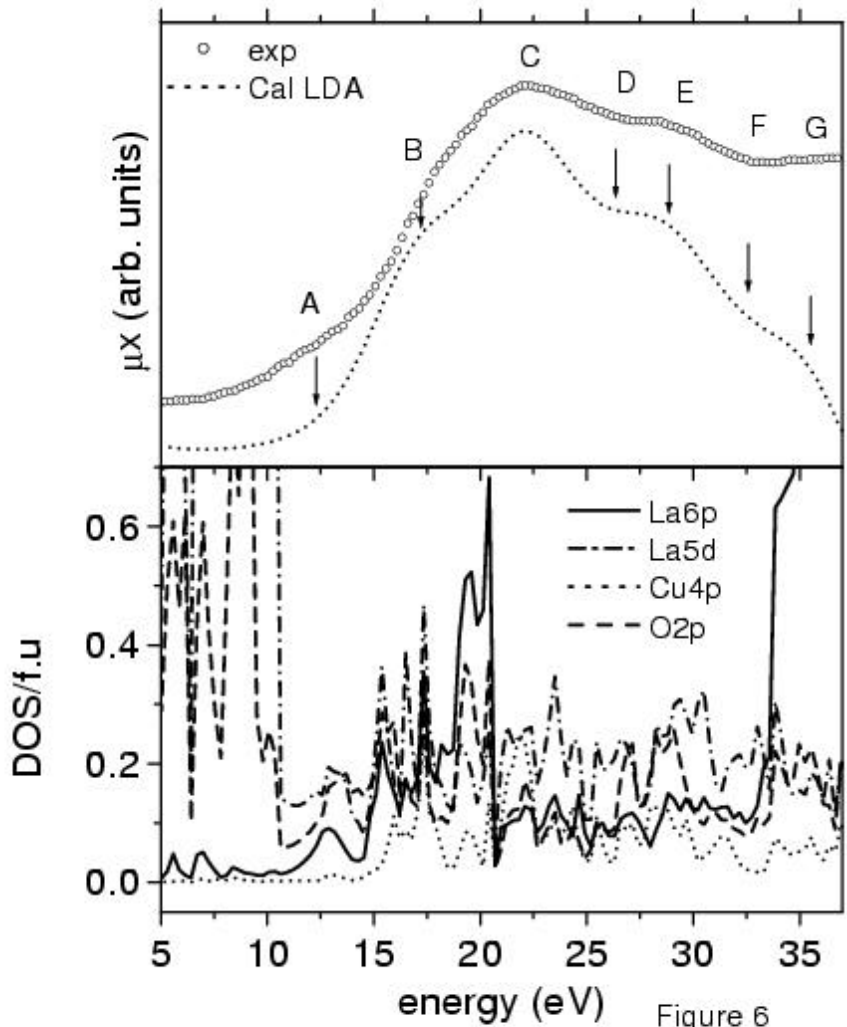

Figure 6
Pandey et al.

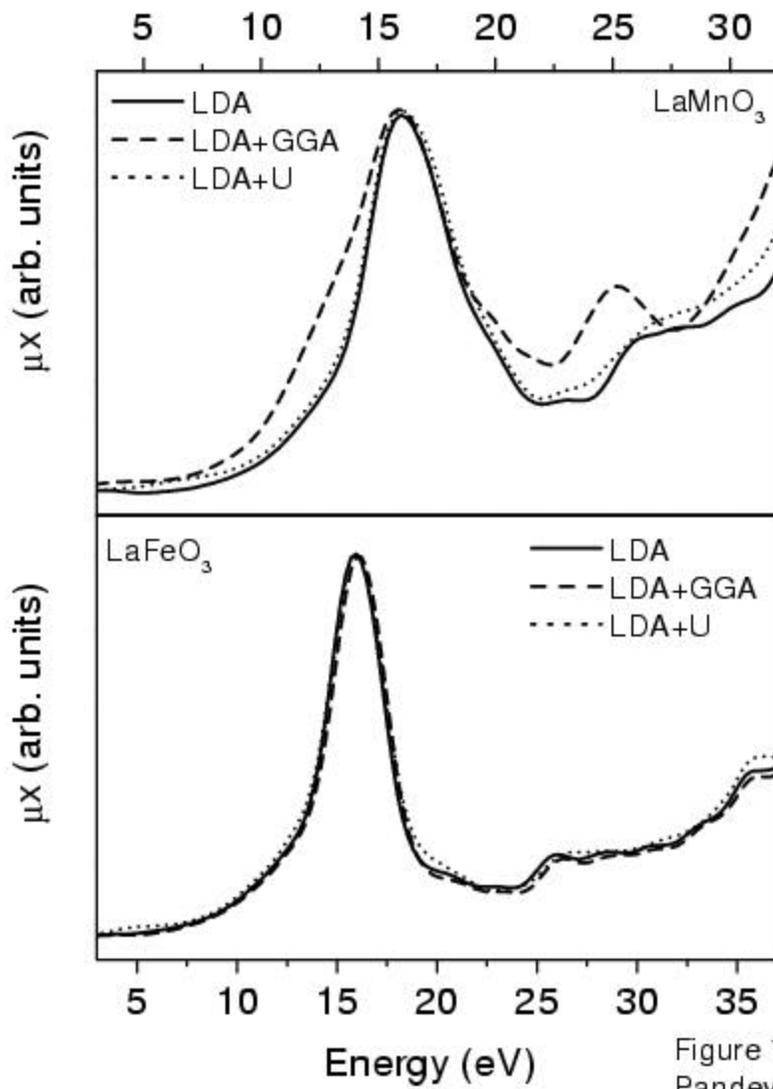

Figure 7
Pandey et al.

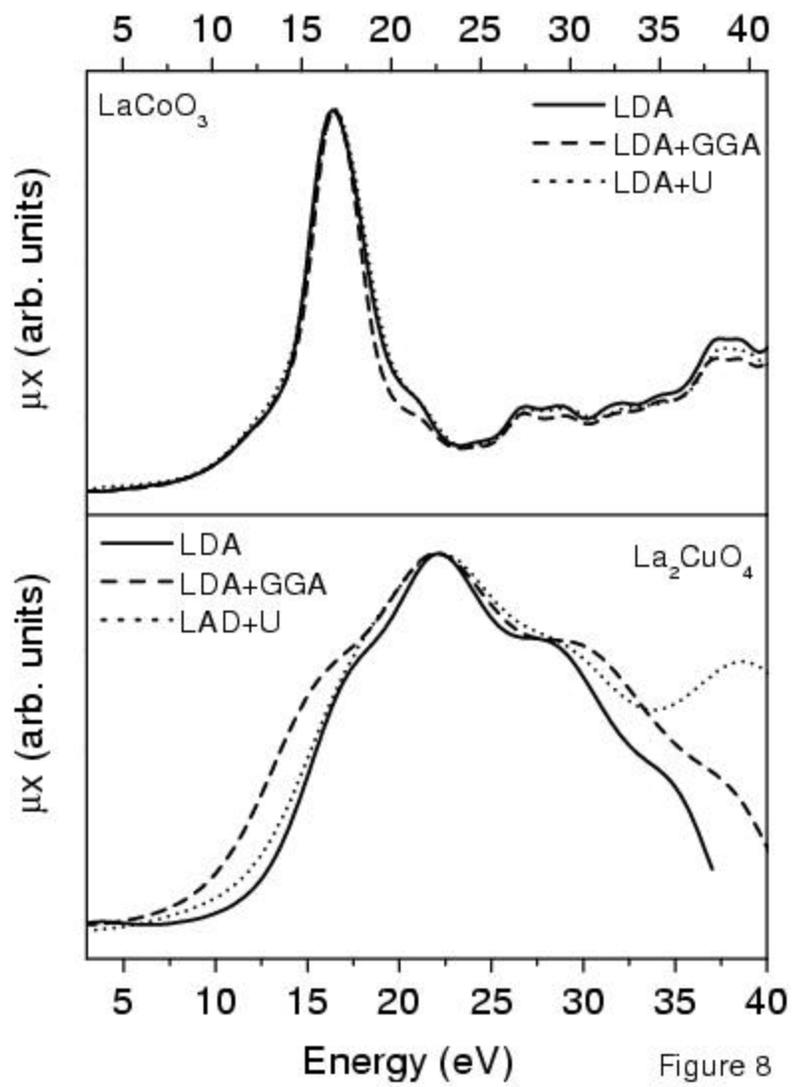

Figure 8
Pandey et al.